\documentclass[twocolumn,superscriptaddress,showpacs,preprintnumbers,amsmath,amssymb,prl]{revtex4}

\usepackage[dvips]{graphicx}
\usepackage{dcolumn}
\usepackage{bm}
\usepackage{amsmath}
\newcommand{\msr}{$\mu$SR}
\newcommand{\lsco}{La$_{2-x}$Sr$_x$CuO$_4$}
\newcommand{\bsco}{Bi$_2$Sr$_2$CuO$_{6+\delta}$}
\newcommand{\bslco}{Bi$_2$Sr$_{2-x}$La$_x$CuO$_{6+\delta}$}
\newcommand{\Tc}{$T_{\rm c}$}
\newcommand{\ns}{$n_{\rm s}$}

\begin{document}

\title{Metal-Insulator Transition and Pseudogap in Bi$_{1.76}$Pb$_{0.35}$Sr$_{1.89}$CuO$_{6+\delta}$ High-$T_c$ Cuprates}

\author{M. Miyazaki}\thanks{Present address: Graduate School of Engineering, Muroran Institute of Technology, Muroran, Hokkaido 050-8585, Japan}
\affiliation{Muon Science Laboratory and Condensed Matter Research Center, Institute of Materials Structure Science, High Energy Accelerator Research Organization (KEK), Tsukuba, Ibaraki 305-0801, Japan}
\author{R. Kadono}\thanks{Corresponding author: ryosuke.kadono@kek.jp}
\affiliation{Muon Science Laboratory and Condensed Matter Research Center, Institute of Materials Structure Science, High Energy Accelerator Research Organization (KEK), Tsukuba, Ibaraki 305-0801, Japan}
\affiliation{Department of Materials Structure Science, Graduate University for Advanced Studies, Tsukuba, Ibaraki 305-0801, Japan}
\author{M. Hiraishi} 
\affiliation{Muon Science Laboratory and Condensed Matter Research Center, Institute of Materials Structure Science, High Energy Accelerator Research Organization (KEK), Tsukuba, Ibaraki 305-0801, Japan}
\author{A. Koda}
\affiliation{Muon Science Laboratory and Condensed Matter Research Center, Institute of Materials Structure Science, High Energy Accelerator Research Organization (KEK), Tsukuba, Ibaraki 305-0801, Japan}
\affiliation{Department of Materials Structure Science, Graduate University for Advanced Studies, Tsukuba, Ibaraki 305-0801, Japan}
\author{K.~M.~Kojima}
\affiliation{Muon Science Laboratory and Condensed Matter Research Center, Institute of Materials Structure Science, High Energy Accelerator Research Organization (KEK), Tsukuba, Ibaraki 305-0801, Japan}
\affiliation{Department of Materials Structure Science, Graduate University for Advanced Studies, Tsukuba, Ibaraki 305-0801, Japan}
\author{Y. Fukunaga}
\affiliation{Department of Applied Physics, Graduate School of Engineering, Tohoku University, Sendai 980-8579, Japan}
\author{Y. Tanabe}\thanks{Present address: Department of Physics, Graduate School of Science, Tohoku University, Sendai 980-8578, Japan}
\affiliation{Department of Applied Physics, Graduate School of Engineering, Tohoku University, Sendai 980-8579, Japan}
\author{T. Adachi}
\affiliation{Department of Engineering and Applied Sciences, Sophia University, Tokyo 102-8554, Japan}
\author{Y. Koike}
\affiliation{Department of Applied Physics, Graduate School of Engineering, Tohoku University, Sendai 980-8579, Japan}

%

\begin{abstract}  
It is inferred from bulk-sensitive muon Knight shift measurement for a  Bi$_{1.76}$Pb$_{0.35}$Sr$_{1.89}$CuO$_{6+\delta}$ single-layer cuprate that metal-insulator (MI) transition (in the low temperature limit, $T\rightarrow0$) occurs at the critical hole concentration $p=p_{\rm MI}=0.09(1)$, where the electronic density of states (DOS) at the Fermi level is reduced to zero by the pseudogap irrespective of the N\'eel order or spin glass magnetism. Superconductivity also appears for $p>p_{\rm MI}$, suggesting that this feature is controlled by the MI transition. More interestingly, 
the magnitude of the DOS reduction induced by the pseudogap remains unchanged over a wide doping range  ($0.1\le p\le0.2$), indicating that the pseudogap remains as a hallmark of the MI transition for $p>p_{\rm MI}$.  
\end{abstract}

\maketitle

The microscopic origin of the pseudogap, or the reduction in the electronic density of states (DOS) observed below a certain onset temperature ($T^*$) in hole-doped high-$T_c$ cuprates, remains elusive despite decades of extensive research.  
From angle-resolved photoemission spectroscopy (ARPES) analysis, it has been inferred that the pseudogap and superconductivity compete with each other and coexist largely by segregating on the Fermi surface, where the carriers primarily resident around the nodes  (comprising the Fermi ``arc") facilitate superconductivity while the pseudogap develops in the antinodal region \cite{ARPES1,ARPES2,2011Kondo}.  
Although there is growing evidence that the pseudogap accompanies certain broken electronic symmetries \cite{Ghiringhelli:12,Chang:12,Comin:14,Neto:14,Hashimoto:15}, it is not clear whether or not these broken symmetries are the origin of the pseudogap.
It is worth remembering that the N\'eel order in lightly doped cuprates is controlled by the inter-layer coupling between the CuO$_2$ layers, which is a material-dependent parameter that is not necessarily relevant to the underlaying energy scale of the intrinsic intra-layer electronic correlation.  In contrast, the metal-insulator (MI) transition (or crossover at finite temperatures) in underdoped cuprates is the direct manifestation of the intra-layer correlation central to the Mott physics, and its relevance to the pseudogap is of crucial importance.


Here, we report on the hole concentration ($p$) dependence of the normal state DOS in Bi$_{1.76}$Pb$_{0.35}$Sr$_{1.89}$CuO$_{6+\delta}$ [(Bi,Pb)2201] which is derived from muon Knight shift ($K_\mu$) measurements under a high transverse field. The physical quantities characterizing the pseudogap, i.e., $T^*$ and the gap energy ($\Delta_1$, assuming $d$-wave symmetry) are also determined from the temperature ($T$) dependence of the shift [$K_\mu(T)$].  We find that the residual shift $K_0$ $[\equiv K_\mu(0)]$ is zero at $p\simeq0.10\equiv p_{\rm MI}$ (the critical concentration), and that $K_0$ develops linearly with $p$. More interestingly, while $T^*$ and $\Delta_1$ exhibit a strong $p$ dependence consistent with earlier reports,  the magnitude of the reduction, $K_{\rm pg}\equiv K_\mu(T^*)-K_0$, demonstrates a least dependence on $p$, indicating that the DOS depleted by the pseudogap is determined by the Fermi surface at $p=p_{\rm MI}$.  Considering the absence of magnetism at $p_{\rm MI}$, these observations suggest that the pseudogap is primarily linked to the momentum-dependent charge localization driven by the intra-layer electronic correlation. 

The (Bi,Pb)2201 compound is a variant of \bsco, in which carrier doping can be attained over a wide range of hole concentrations, i.e., from lightly doped ($p\le0.1$) to overdoped ($p\ge 0.2$) by controlling the oxygen content $\delta$.  In contrast to \lsco\ (LSCO), the strong intra-layer antiferromagnetic (AF) correlation does not lead to instability of the spin glass or the N\'eel order in the lightly doped region; this behavior most likely results from the large distance between the CuO$_2$ layers ($\simeq1.22$ nm, almost twice as large as that of LSCO) \cite{Russo,Bi2201ZFmSR,Enoki:2013}. 
This feature provides a considerable advantage for muon spin rotation (\msr), because the majority of high-$T_{\rm c}$ cuprates exhibit the N\'eel order in the relevant doping range which precludes high precision frequency shift measurements for investigating the DOS \cite{LSCOIshida}. Moreover, the (Bi,Pb)2201 compound exhibits relatively low $T_c$ ($\le20$ K) and a correspondingly low irreversibility field ($B_{\rm irr}<6$ T); thus, it is feasible to study the normal state DOS below $T_c$ by suppressing the superconducting gap under a modest external field \cite{Hc2}. 
The samples examined in this study and their bulk properties are summarized in the Supplemental Material \cite{Suppl} (see also Fig.~\ref{kmu_T} inset for a quick reference to their label and $T_c$ vs $p$).  
 
A conventional \msr\ experiment was conducted on the TRIUMF M15 beamline using the HiTime spectrometer. An external field of 6 T ($B_0,\:\parallel \hat{z}$ axis) was applied parallel to the $c$-axes of the (Bi,Pb)2201 crystals for all Knight shift measurements, where the field was sufficiently high to suppress the superconductivity of all the investigated samples. The complex decay positron asymmetry [$A(t)=A_x(t)+iA_y(t)$] was monitored by two pairs of scintillation counters placed along the $\hat{x}$ and $\hat{y}$ directions. The sample $T$ was controlled over a $T$ range of 2 to 300 K using a helium gas flow cryostat.   

\begin{figure}[tb]
	\centering
	\includegraphics[width=0.45\textwidth,clip]{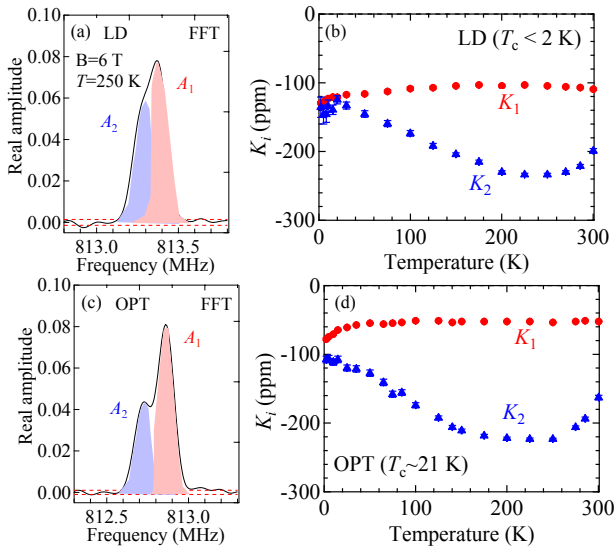}
	\caption{(Color online)  Examples of fast Fourier-transformed \msr\ spectra observed at $\sim$250 K in (a) LD and (c) OPT, with the temperature dependence of the muon Knight shift $K_i$ ($i=1,2$, corresponding to the signal with amplitude $A_i$)  in the respective samples being shown in (b) and (d). The small downturn below $\sim$50 K in (d) is commonly observed for $K_1$ and $K_2$  and is due to the diamagnetism of the impurities (see text).
	 }
	\label{spectra}
\end{figure}

 Typical examples of fast Fourier-transformed \msr\ time spectra observed at 250 K for the LD and OPT samples are shown in Figs.~\ref{spectra}(a) and (c), where two lines with comparable amplitudes are clearly discernible. This feature is apparent for all examined samples, and their relative amplitude is primarily sample independent; this result strongly suggests that the signal splitting is due to muons occupying two inequivalent sites with different hyperfine parameters in the unit cell (see below).  This conclusion is also in line with the result of a previous ZF-\msr\ study on the AF phase of La-Bi2201, which suggests the presence of two different sites for muons probing different internal magnetic fields \cite{Labi2201_zfmsr}. 

The \msr\ spectra were analyzed in the time domain using conventional least-squares fitting with appropriate modeling of the two signal components, with
\begin{equation}\label{eq:analysis}
A(t)\simeq A_1e^{-\lambda_1t} e^{i(2\pi f_1t+\phi)}+A_2e^{-\sigma_2^2t^2}e^{i(2\pi f_2t+\phi)},
\end{equation}
where $A_i$ indicate the initial amplitudes ($i=1,2$),  $\lambda_1$ and $\sigma_2$ are the depolarization rates, $f_i$ are the precession frequencies, and $\phi$ is the initial phase of precession. The line shape for depolarization (either  exponential or Gaussian) was chosen so as to minimize the chi-square values for the respective components after several trials. Because of the small depolarization rates, $A_i$ were best determined at 250 K for LD--OPT and at 225 K for OD--NSOD, exhibiting a relative amplitudes $A_1$:$A_2\simeq$ 6:4--7:3 for all samples.  This supports the scenario of the two inequivalent muon sites. $A_1$:$A_2$ was fixed to the high $T$ value for the remainder of the analysis, so that other parameters could be extracted via curve fitting.

Examples of the frequency shift, $K_i=(f_i-f_0)/f_0$, versus $T$ are shown in Figs.~\ref{spectra}(b) and (d) for the respective components ($i=1,2$), where $f_0=\gamma_\mu B_0/2\pi$ ($\gamma_\mu=2\pi\times 135.53$ MHz/T is the muon gyromagnetic ratio) was determined from reference data on a silver plate (to correct the slight variance of $f_0$ against $T$ due to thermal contraction of the cryostat relative to the magnet, etc.) as $f_0=f_{\rm Ag}/(1+K_{\rm Ag})$, with $f_{\rm Ag}$ and $K_{\rm Ag}$ being the muon precession frequency and corresponding Knight shift, respectively.  Here, we adopted $K_{\rm Ag} = 94$ ppm \cite{KAg}. 

The behavior of $K_i$ below $\sim$220 K in LD and $\sim$180 K in OPT is consistent with the reduction of the local spin susceptibility accompanied by the pseudogap, whereas the turnover of $K_i$ commonly observed above $\sim$250 K suggests the occurrence of an extrinsic effect (e.g., muon diffusion) induced by thermal activation.  The small downturn of $K_i$ below $\sim$50 K in Fig.~\ref{spectra}(d) is attributed to the Lorentz demagnetization ($= -N_z\chi_0B_0/4\pi$, with $N_z\simeq1$ for the present thin-sheet samples), where the uniform susceptibility exhibited a Curie term ($\chi_0\propto1/T$) that was most likely due to impurities \cite{Bi2201ZFmSR}.  
Apart from these extrinsic features, the direction and magnitude of the change in $K_i$ imply that the hyperfine parameter $A_{\mu(1)}$ ($A_{\mu(2)}$) for $K_1$ ($K_2$) is positive (negative), with $|A_{\mu(1)}|$ being considerably smaller than $|A_{\mu(2)}|$. This finding is perfectly in line with the prediction for the muon sites estimated using the VASP code \cite{VASP}, i.e., $A_{\mu(1)}=+0.065$ T/$\mu_B$ for the site near the BiO layer and  $A_{\mu(2)}=-0.337$ T/$\mu_B$ for that in the CuO$_2$ plane (see Supplemental Material \cite{Suppl}).

In general, $K_i$ consists of multiple contributions 
\begin{equation}
K_i=K_{\mu(i)}+K_{\rm dia}+K_{\rm inst}, \:\:(i=1,2)\label{eq:Kmu}
\end{equation}
 where $K_{\mu(i)}$ is the spin part of the muon Knight shift, $K_{\rm dia}$ is the diamagnetic correction including the Lorentz demagnetization, and $K_{\rm inst}$ is an instrumental offset.  The spin part of the muon Knight shift is proportional to the static local spin susceptibility ($\chi_{\rm loc}$) at the muon site, 
and they are directly related to the DOS at the Fermi level $D(E_F)$, as 
\begin{equation} 
K_{\mu(i)}\simeq A_{\mu(i)}\frac{\chi_{\rm loc}}{N_A\mu_{\rm B}}=A_{\mu(i)}\frac{2\mu_{\rm B}}{N_A}D(E_F),\label{eq:Kmu_DOS}
\end{equation}
where  $\mu_{\rm B}$ is the Bohr magneton and $N_{\rm A}$ is the Avogadro number.  Considering the fact that $A_{\mu(1)}\cdot A_{\mu(2)}<0$, Figs.~1(c) and 1(d) indicate that $K_{\rm inst}\simeq-100$ppm. Thus, it is usually difficult to correct $K_{\rm dia}$ and $K_{\rm inst}$ precisely for every sample. Fortunately, these extrinsic contributions are independent of the muon sites, and they can be eliminated by taking differential values between $K_1$ and $K_2$, i.e.,
\begin{equation}
 \overline{K}_\mu=K_1-K_2 = K_{\mu(1)} -K_{\mu(2)}\equiv cK_\mu,
\end{equation}
 where $c$ is the numerical factor used to link the true $K_\mu$ with the observed shifts regarding the spin part. (For the individual values of $K_i(T)$, see the supplemental material \cite{Suppl}.)
It should be stressed that, owing to the fortunate scenario involving two muon sites with different hyperfine parameters, $ \overline{K}_\mu$ is free from the ambiguity of various offsets. Thus, the null shift was determined from the crossing point of $K_1(T)$ and $K_2(T)$, with $T$ as an internal parameter. For the quantitative discussion, $\Delta_1$ was estimated from curve fits of the data for $T\le T^*$ shown in Fig.~\ref{kmu_T} assuming a $d$-wave gap, 
\begin{equation}\label{eq:dW}
\overline{K}_\mu=K_0+K_{\rm pg}D(E_F:\Delta_1),
\end{equation} 
where $K_0$ and $K_{\rm pg}$ are free parameters, $D(E_F:\Delta_1)=\frac{1}{k_BT}\int\int \frac{d\phi}{2\pi} d\epsilon\cos^2\phi \cosh^{-2}[\sqrt{\epsilon^2+[\Delta_1(T)\cos2\phi]^2}/2k_BT]$ with $\Delta_1(T)$ being the BCS gap \cite{Prozorov:08}.   $T^*$ is defined as the temperature at which $\overline{K}_\mu$ decreases from $K_{\rm max}$ by the width of the error bar.  The curves obtained from the least-square minimization exhibit reasonable agreement with the data (except for $T>250$ K, where the influence of muon diffusion is evident).  As can be seen from Fig.~\ref{kmu_T}, the parameter errors are dominated by the statistical precision of $\overline{K}_\mu$.

\begin{figure}[tb]
	\centering
	\includegraphics[width=0.48\textwidth]{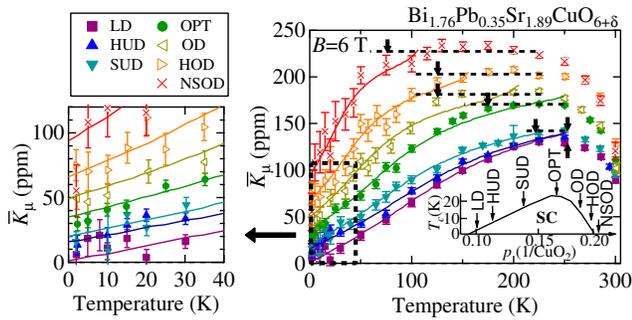}
	\caption{(Color online) Temperature dependences of muon Knight shift $\overline{K}_\mu$ (at 6 T) in (Bi,Pb)2201.  A magnified view is shown for $T<40$ K (left). The solid curves are fits obtained using Eq.~(\ref{eq:dW}).  The onset temperature of the pseudogap ($T^*$) is indicated by black arrows. Inset: \Tc\ vs.~$p$ for measured samples.} 
	\label{kmu_T}
\end{figure}

 \begin{figure}[b]
	\centering
	\includegraphics[width=0.47\textwidth]{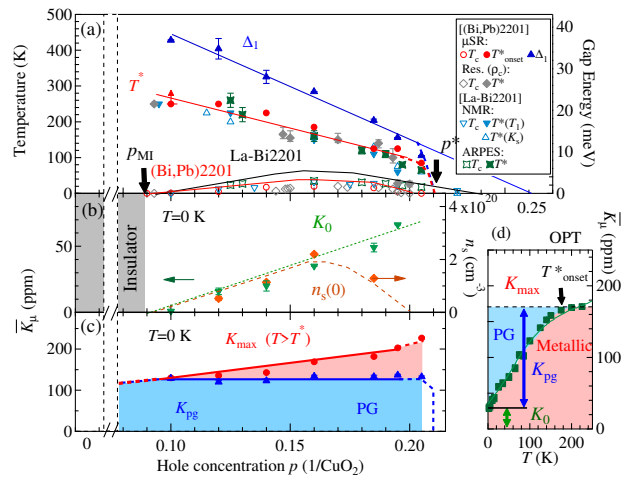}
	\caption{(Color online)  (a) Hole concentration ($p$) dependence of  \Tc, $T^*$, and $\Delta_1$ in (Bi,Pb)2201. The $c$-axis resistivity data ($\rho_{\rm c}$) are quoted from Refs.\cite{Kudo_PG1,Kudo_PG2,Kudo_Tc}, the NMR data are from Refs.~\cite{NMR1, NMR2}, and the ARPES data are from Ref.~\cite{2011Kondo}.  (b) Superfluid density [\ns(0), brown rhombuses] and $K_0$ (inverted green triangles) vs.~$p$, where the latter corresponds to the residual DOS for $T\rightarrow 0$ in the normal state. (c) $K_{\rm max}$ (red squares, corresponding to DOS at $T>T^*$) and $K_{\rm pg}$ (blue triangles,  corresponding to DOS depleted by pseudogap) vs.~$p$. (d) Definitions of $K_{\rm max}$, $K_{\rm pg}$, and $K_0$ for $\overline{K}_\mu$ in OPT (quoted from Fig.~\ref{kmu_T}).  
 }
	\label{fig:phase-dgm}
\end{figure}
Figure \ref{fig:phase-dgm}(a) shows  $T^*$ vs. $p$, which is in reasonable agreement with the results inferred from a variety of analyses, including 1/$T_1$ and Knight shift from nuclear magnetic resonance (NMR), and ARPES on La-Bi2201 \cite{NMR1,NMR2,ARPES1},  and the $c$-axis resistivity ($\rho_{\rm c}$) on oxygen-controlled (Bi,Pb)2201 \cite{Kudo_PG1,Kudo_PG2}. Thus, the pseudogap phenomenon is confirmed by the muon Knight shift as regards the behavior of $T^*$.  It may be worth mentioning that  NSOD ($T_{\rm c}=0$ K) exhibits a finite $T^*$ and $\Delta_1$, but no such behavior was indicated by a previous NMR study of an overdoped La-Bi2201 sample ($T_{\rm c}=8$ K) \cite{NMR2}. 
The magnitude of $\Delta_1$ inferred from \msr\ analysis for $p<0.2$ is also consistent with that inferred from ARPES and other investigations of various cuprates that exhibit a monotonous decrease with $p$, being linearly extrapolated to zero at $p=0.25(5)$ \cite{ARPES2,EPG,Bi2212STM,LSCO_PES,Miyakawa}. However, the magnitude of $\Delta_1$ exhibits a steep decrease for $p\ge0.2$, and it disappears at a $p$ slightly lower than that of the other single-layer cuprates.
In any case, it may be argued that the pseudogap disappears around 0.2$<p<$0.22 [$p^*=0.21(1)$] in Bi2201, regardless of the differences in the hole doping method and the associated variation of \Tc.

\begin{figure}[tb]
	\centering
	\includegraphics[width=0.4\textwidth]{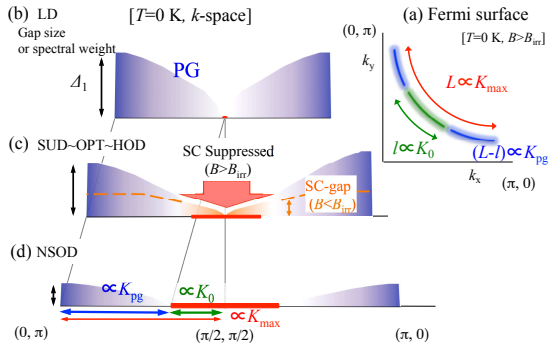}
	\caption{(Color online) (a) Schematic illustration of correspondence between $K_{\rm max}$,  $K_{\rm pg}$,  $K_0$, and DOS at Fermi level in momentum space. (b) $K_{\rm max}\simeq K_{\rm pg}$ suggests a point-like residual DOS ($K_0\simeq0$) at $p=p_{\rm MI}=0.09(1)$. (c) The increase of $K_0$ with $p$ suggests the development of DOS around the nodes. (d) Continuous increase of DOS with $p$ as \Tc\ is reduced to zero. The magnitude of the pseudogap ($\Delta_1$) decreases monotonously with increasing $p$.}\label{fig:k-space} 
\end{figure}

The behavior of  $K_{\rm max}$, $K_{\rm pg}$, and $K_0$  may be understood using a model in the momentum space suggested by the ARPES analysis.  As $D(E_F)$ ($\propto[dE({\bm k})/d{\bm k}_F]^{-1}$) is thought to be independent of the Fermi momentum (${\bm k}_F$) in Bi2201 \cite{ARPES_hashi}, $K_{\rm max}$ is proportional to the entire DOS at high temperatures, $D(E_F,T^*)$. Therefore, $K_{\rm max}$ corresponds to the high-temperature Fermi surface [$\propto L$ in Fig.~\ref{fig:k-space}(a)]. The increase of $K_{\rm max}$ in proportion to $p$ indicates the evolution of the Fermi surface caused by the shift of the chemical potential ($=E_F$), which is consistent with the rigid-band behavior suggested by the results of ARPES analyses \cite{ARPES_hashi}. Meanwhile, $K_0$ corresponds to the remaining DOS at $T\rightarrow0$, being proportional to the residual length of the Fermi arc ($l$). The curve fit obtained for the relation $K_0\propto(p-p_{\rm MI})$ yields  $p_{\rm MI}=0.09(1)$. (Note that the superconducting gap is suppressed by an external field greater than $B_{\rm irr}$.)  Consequently, $K_{\rm pg}$ is a measure of the magnitude of the depleted DOS upon the opening of the pseudogap ($K_{\rm pg}\propto L-l$).
Thus, the fact that $K_{\rm pg}$ is mostly independent of $p$ [see Fig.~\ref{fig:phase-dgm}(c)] indicates that a fixed amount of DOS (determined by $p_{\rm MI}$) is depleted by the pseudogap irrespective of $p$. This finding also suggests that the metal-insulator crossover occurs when $K_{\rm pg}=K_{\rm max}$, which in turn determines $p_{\rm MI}$.

From a previous Cu-NMR study, it has been inferred that the N\'eel order occurs at $p=p_{\rm N}\simeq0.10$--0.11 in La-substituted Bi2201 (\bslco, $T_{\rm N}\simeq50$ K) \cite{NMR2}, while the ARPES data suggests that $p_{\rm MI}\simeq0.07$--0.08 \cite{ARPES_hashi}. This implies that \bslco\ has a relatively stronger inter-layer coupling.  This leads to the increase of $T_{\rm N}$ and $p_{\rm N}$ in comparison with (Bi,Pb)2201, where $p_{\rm MI}$ ($<p_{\rm N}$) is untraceable by the Cu-Knight shift because of the N\'eel order. This also provides a natural explanation for the presence of residual DOS at $p_{\rm N}$ ($>p_{\rm MI}$), which has been inferred from ARPES results. In contrast, the present (Bi,Pb)2201 system remains nonmagnetic (there is no indication of the N\'eel order or spin glass magnetism) at $p=0.09$ \cite{LD1}, which allows clear identification of $p_{\rm MI}$ based on the muon Knight shift.   Meanwhile, the NMR shift corresponding to $K_{\rm pg}$ has been confirmed to be primarily independent of $p$, at least for $p> p_{\rm N}$ \cite{Kawasaki}, which is perfectly in line with the present result.  Thus, it is strongly suggested that $p_{\rm MI}$ and, hence, the MI transition at $T=0$, are uniquely correlated with the pseudogap, irrespective of the N\'eel order. A similar correlation is suggested for LSCO, as the residual DOS approaches zero for $p<0.05$ \cite{Shen:05}, where the MI transition is inferred from the differential in-plane resistivity (DIPR) \cite{Ando:04}. The present result provides circumstantial microscopic evidence for the empirical doping phase diagram derived from the DIPR measurements  \cite{Ando:04}.

Now, let us discuss the relationship between the pseudogap and superconductivity. 
Fig.~\ref{fig:phase-dgm}(b) shows the superfluid density at $T\rightarrow0$ [$n_{\rm s}(0)$] quoted from our previous report on (Bi,Pb)2201 \cite{2007Miyazaki}, where we assumed that the effective carrier mass was 3--4$m_{\rm e}$ and almost independent of $p$ \cite{Russo,eff_mass}.   While $n_{\rm s}(0)$ exhibits a dome-like $p$ dependence similar to that of \Tc, it does not exhibit any clear correlation with other quantities related to the pseudogap ($T^*$, $\Delta_1$, and $K_{\rm pg}$). 

In contrast, $n_{\rm s}(0)$ seems to be proportional to $K_0$ in the underdoped region ($p\le0.16$), suggesting that  the majority of the mobile holes ($\simeq p-p_{\rm MI}$) are involved in the Cooper pair formation. This hints at the correspondence between the MI crossover and superconductor-insulator transition observed in one-unit-cell thick \lsco\ \cite{Bollinger:11}.  [The decrease of $n_{\rm s}(0)$ against $K_0$ for $p>0.16$ suggests the occurrence of electronic phase separation in the overdoped region.]   The fact that $T_c\propto n_{\rm s}(0)$ in the underdoped region is naturally understood by acknowledging that the superconducting gap [$\Delta({\bm k}_F)$] is maximal at approximately the edge of the Fermi arc for the $d$-wave pairing \cite{Matsuzaki:04}, i.e.,
$\Delta_{\rm max}({\bm k}_F)\propto K_0\propto l$, which is perfectly in line with the earlier suggestion from the ARPES analyses \cite{ARPES2}.

In summary, it was inferred from muon Knight shift measurements in (Bi,Pb)2201 that the metal-insulator crossover (or transition at $T=0$) occurs at $p=p_{\rm MI}\simeq0.09(1)$ and does not accompany the magnetic instability. This finding, together with the previous Cu-NMR result for \bslco, suggests that the MI transition and pseudogap are closely related features relevant to the instability of the charge sector in cuprates.


 We would like to thank the TRIUMF staff for their technical support during the \msr\ experiment. 
 We would also like to express our gratitude to H. Kumigashira, R. Yukawa, H. Nakao, K. Yamada, and S. Kawasaki for helpful discussion.  This work was partially supported by the KEK-MSL Inter-University Program for Overseas Muon Facilities and a Grant-in-Aid for Scientific Research on Priority Areas (Grant No.~16076211) from the Ministry of Education, Culture, Sports, Science and Technology, Japan.

\end{document}